\documentclass[10pt,a4paper]{amsart}
\usepackage{amsmath,amsthm,amscd,amssymb,times}

\def\C{\mathbb{C}}

\def\Q{\mathbb{Q}}

\def\eps{\varepsilon}

\newcommand{\be}{\begin{equation}}
\newcommand{\ee}{\end{equation}}
\newcommand{\bea}{\begin{eqnarray}}
\newcommand{\eea}{\end{eqnarray}}
\newcommand{\beas}{\begin{eqnarray*}}
\newcommand{\eeas}{\end{eqnarray*}}

\newtheorem{theorem}{Theorem}

\newtheorem{prop}[theorem]{Proposition}

\begin{document}

\title[The residues of quantum field theory - numbers we should know]{The residues of quantum field theory - numbers we should know${}^\dagger$}
\author[Dirk Kreimer]{Dirk Kreimer${}^\ast$}
\address{Institut des Hautes \'Etudes Scientifiques, 35 rte.~de Chartres, F-91440 Bures-sur-Yvette \indent and
Center for Mathematical Physics, Boston University, Boston MA 02215.} \email{kreimer@ihes.fr} \thanks{\hspace{-5mm}${}^\ast$CNRS at IHES.\\
${}^\dagger$Contributed to the proceedings of the {\em Workshop on Noncommutative Geometry and Number Theory}, August 18-22 2003, Max Planck Institut f\"ur
Mathematik, Bonn. [hep-th/0404090]}

\begin{abstract}
We discuss in an introductory manner structural similarities between the polylogarithm and Green functions in quantum field theory.
\end{abstract}

\maketitle

\section{Introduction: ambiguities in the choice of either a branch or a finite part}
It is a pleasure to report here on a connection between mathematics and physics through the study of Dyson--Schwinger equations (DSE) which has been left mostly
unexplored so far. While a thorough study of these quantum equations of motions for four-dimensionable renormalizable gauge field theories is to be presented in
\cite{new}, here we have a much more limited goal: to introduce this connection in simple examples and use it as a pedagogical device to explain how the Hopf
algebraic structure of a perturbative expansion in quantum field theory (QFT), those non-perturbative quantum equations of motion, renormalization and (breaking
of) scaling behaviour fit together.

\subsection{The polylog} We will start our exploration in the rather distinguished world of polylogarithms and mixed Tate Hodge structures to have examples for such phenomena.
We emphasize right away though that non-trivial algebraic geometry considerations are beyond our scope. If the remarks below familiarize the reader with this
very basic connection between the structure of quantum field theory and such objects they have fulfilled their goal.

Consider the following $N\times N$ matrix $M^{(N)}$ borrowed from Spencer Bloch's function theory of the polylogarithm \cite{Spencer1} (see also \cite{Oest} and
references there): \be
\begin{array}{c}
  \alpha^0 \\
  \alpha^1 \\
  \alpha^2 \\
  \alpha^3 \\
  \ldots
\end{array}
\left(
\begin{array}{rcrcrcrcr}
  + 1 & \mid & 0 & \mid & 0 & \mid & 0 & \mid & \cdots\\
  -{\rm Li}_1(z) & \mid & 2\pi i & \mid & 0 & \mid & 0 & \mid & \cdots \\
  -{\rm Li}_2(z)  & \mid & 2\pi i \ln z & \mid & [2\pi i]^2 & \mid & 0 & \mid &  \cdots\\
  -{\rm Li}_3(z)  & \mid & 2\pi i \frac{\ln^2 z}{2!} & \mid & [2\pi i]^2\ln z & \mid & [2\pi i]^3 & \mid &  \cdots\\
  \ldots & \mid & \ldots & \mid & \ldots & \mid & \ldots & \mid & \ldots
\end{array}\right),
\ee here spelled out for $N=4$. Note that we assign an order in a small parameter $\alpha$ to each row, counting rows $0,1,\ldots$ from top to bottom, similarly
we count columns $0,1,\ldots$ from left to right. We use the polylog defined by \be {\rm Li}_n(z)=\sum_{k=1}^\infty\frac{z^k}{k^n} \ee inside the unit circle and
analytically continued with a branch cut along the real axis from one to plus infinity, say.

The matrix above is highly structured in that the ambiguity reflected by the branch cut, for any entry $M_{i,j}$, is nicely stored in the same row $i$ at
$i,j+1$. Furthermore in each column from top (disregarding the trivial uppermost row $1,0,0,\ldots$) to bottom each transcendental function ${\rm Li}_n(z)$ or
$\ln^m(z)/m!$ has the same coefficient: $-1$ in the first column, $2\pi i$ in the second, and so forth. This structure allows for the construction of unambiguous
univalent polylogs \cite{Spencer1} assembled from real and imaginary parts of those rows, for example the univalent dilog is $ \Im({\rm Li}_2(z))+\ln\mid\!
z\!\mid \arg(1-z) $.

\subsection{DSE for the polylog} With this motivic object thrown at us, we can familiarize ourselves with it by considering the following Dyson--Schwinger equation, where the use of
this name is justified from the basic observation that it can be written using the Hochschild cohomology of a Hopf algebra of the underlying perturbative
expansion \cite{annals,Kreimer2} as exemplified below. Consider, for suitable $z$ off the cut, \be
F(\alpha,z)=1-\frac{1}{1-z}+\alpha\int_0^z\frac{F(\alpha,x)}{x}dx,\ee where we continue to name-drop as follows: We call $F(\alpha,z)$ a renormalized Green
function, $\alpha$ the coupling (a small parameter, $0<\alpha<1$) and consider the perturbative expansion \be F(\alpha,z) =1-\frac{1}{1-z}+\sum_{k=1}^\infty
\alpha^k f_k(z),\ee where we distinguished the lowest order term $f_0(z)=z/(z-1)$ (which corresponds to the term without quantum corections in QFT) at order
$\alpha^0$ which here equals $-{\rm Li}_0(z)$, rather consistently. We immediately find \be f_1(z)=\ln(1-z)=-{\rm Li}_1(z)\ee and if we remind ourselves that the
log is a multivalued function with ambiguity an integer multiple of $2\pi i$, we reproduce the second row in the above. Identifying row numbers with powers of
$\alpha$ increasing from top to bottom the above matrix does then nothing but providing the solution of the DSE so constructed: \be \sum_{j\leq
k}M_{k,j}=f_k(z),\;k>0.\ee

We now utilize the Hopf algebra $H$ of non-planar undecorated rooted trees \cite{CK1}. It has a Hochschild 1-cocycle $B_+:\;H\to H$, such that it determines the
coproduct $\Delta$ via the closedness of this cocycle, \be bB_+=0 \Leftrightarrow \Delta B_+=B_+\otimes 1 +[{\rm id}\otimes B_+]\Delta\ee and $\Delta(1)=1\otimes
1$.

There is a sub-Hopf algebra of "ladder trees" \be t_n:=\underbrace{B_+(B_+(\cdots(B_+(1))\cdots)}_{n-{\rm times}}.\ee For them, we have \be
\Delta(t_n)=\sum_{j=0}^n t_j\otimes t_{n-j},\ee which is cocommutative and we identify $t_0=1_H$. For these ladder trees $t_n$ we also introduce an extra
dedicated commutative product $t_n\cdot t_m=\frac{(n+m)!}{n!m!}t_{n+m}.$ In general, the commutative product in the Hopf algebra $H$ is the disjoint union of
trees into forests \cite{CK1}.

We now define Feynman rules as characters on the Hopf algebra. It thus suffices to give them on the generators $t_n$. Also, as $H$ decomposes as $H=1_H\C\oplus
H_{\rm aug}$, each $h\in H$ decomposes as $h=h_1 + h_{\rm aug}$. We now define our Feynman rules by \be \phi(B_+(h))(z,z_0)=\int_{z_0}^z \frac{\phi(h_{\rm aug
})(x,z_0)}{x}dx+\int_{z_0}^z \frac{\phi(h_1)(x,z_0)}{x-1}dx,\;\forall h\in H,\ee while we set $\phi(1_H)(z,z_0)=1$, and $\phi(h_1h_2)=\phi(h_1)\phi(h_2),$ as
they are elements of the character group of the Hopf algebra. Note that for $X=t_{n+1}=B_+(t_n)$, this gives iterated integrals.

Next, we introduce the series \be H[[\alpha]]\ni X\equiv c_1+\sum_{k=1}^\infty x_k \alpha^k= c_11_H+\alpha B_+\left(\frac{1}{c_1}\bar{e}(X)+P(X)\right),\ee
$c_1=1-\frac{1}{1-z}$ fixing the inhomogenous part. Here, $P$ is the projector into the augmentation ideal, and $\bar{e}$ the counit. Solving this fix-point
equation determines \be X=c_1+\alpha B_+(1)+\alpha^2B_+(B_+(1))+\ldots,\ee hence $x_k=t_k$, $k>0$. We then have $\forall k\geq 0$ \be f_k(z)=\phi(t_k)(z,0),\ee
and the Hochschild closed 1-cocycle $B_+$ maps to an integral operator $\phi(B_+)\to \int dx/x$, as one expects from \cite{CK1}.

Now, let $Li\equiv Li(z)$ and $L\equiv L(z)$ be the characters on the Hopf algebra defined by \be -\phi(t_n)(z,0)\equiv Li(t_n)(z)={\rm
Li}_n(z),\;L(t_n)(z)=\frac{\ln^n(z)}{n!}.\ee From \cite{Spencer1} we know that the elimination of all ambiguities due to a choice of branch lies in the
construction of functions $a_p(z)=(2\pi i)^{-p}\;\widetilde{a}_p(z)$ where \be \widetilde{a}_p(z):=\left[ {\rm Li}_p(z)-\cdots+ (-1)^j {\rm
Li}_{p-j}(z)\frac{\ln^j(z)}{j!}+\cdots +(-1)^{p-1}{\rm Li}_1(z)\frac{\ln^{p-1}(z)}{(p-1)!}\right].\ee We have
\begin{prop} For $z\in\C$, \be \tilde{a}_p(z)=m\circ ((L^{-1}\otimes Li)\circ({\rm id}\otimes P)\circ\Delta(t_p),\ee where $L^{-1}=L\circ S$, with $S$ the antipode in $H$ and $P$ the
projection into the augmentation ideal.
\end{prop}
Proof: elementary combinatorics confirming that $L\circ S(t_n/n!)=(-\ln(z))^n/n!$.

\medskip

 There is a strong analogy here to the Bogoliubov $R$ operation in renormalization theory \cite{annals,int2}, thanks to the fact that $Li$ and $L$
have matching asymptotic behaviour for $\mid\! z\!\mid\to\infty$. Indeed, if we let $R$ be defined to map the character $Li$ to the character $L$, $R(Li)=L$, and
$P$ the projector into the augmentation ideal of $H$, then \be L\circ S=S_R^{Li}=-R[m\circ(S^{Li}_R\otimes Li)({\rm id}\otimes P)\Delta]\equiv
-R\left[\overline{Li}\right], \ee for example \be S_R^{Li}(t_2)=-R[Li(t_2)+S_R^{Li}(t_1)Li(t_1)]=-L(t_2)+L(t_1)L(t_1)=\frac{+\ln^2(z)}{2!},\ee where
$\overline{Li}(t_2)=Li(t_2)-L(t_1)Li(t_1)$. Thus, $a_p$ is the result of the Bogoliubov map \be\overline{Li}=m(S^{Li}_R\otimes Li)({\rm id}\otimes
P)\Delta\label{prop1}\ee acting on $t_n$. The notions of quantum field theory and polylogs are close indeed.

Let us us now reconsider the above function $F(\alpha,z)$ as a function of the lower boundary as well: \be F(\alpha,z)\equiv F(\alpha,z,0)\ee and let us return
to a generic lower boundary ($\not= 1$, say) $z_0$, with corresponding DSE \be F(\alpha,z,z_0)=1-\frac{1}{1-z}+\alpha\int_{z_0}^z \frac{F(\alpha,x,z_0)}{x}dx,\ee
and returning to Feynman characters (for $h\in H_{\rm aug}$) \be \phi(B_+(h))(z,z_0)=\int_{z_0}^z \frac{\phi(h)(x,z_0)}{x}dx.\ee How can we express
$F(\alpha,z,z_0)$ in terms of characters $\phi(z,\tilde{z_0})$ and $\phi(z_0,\tilde{z_0})$?

The answer is given by reminding ourselves that along with the Hopf algebra structure comes the convolution \be
\phi(z,z_0)=m\circ(\phi(z_0,\tilde{z_0})\otimes\phi(z,\tilde{z_0}))\circ (S\otimes{\rm id})\circ\Delta,\ee which answers this question. This is a first example
of renormalization, aimed at a repa\-ra\-metrization in the DSE. Note that here it is understood that maps like $\phi(z,z_0)$ are characters on the Hopf algebra:
$\phi(z,z_0): H\to\C$, when evaluated on a Hopf algebra element for fixed $z,z_0$. Evaluated on an element $t_n$, they reproduce the corresponding element in the
expansion to order $\alpha^n$ of $F(\alpha,z,z_0)$.

\section{Renormalization vs Polylogs}
Having made first contact with renormalization as a modification of a boundary condition in a DSE, we now investigate its greatest strength: the definition of
locality and the absorption of short-distance singularities. To do so, we start with examples which are even simpler than the polylog. So let us now introduce a
first toy model for renormalization still in analogy with the previous section.

\subsection{The simplest model: $F(\alpha,z)=z^{-{\rm Res}(\wp)\alpha}$}
To make close contact with the situation in perturbative quantum field theory we introduce a regulator $\eps$, which is a complex parameter with small positive
real part. For fixed $0<\alpha<1$ we then consider the following equation: \be F_Z(\alpha,z;\eps)=Z+\alpha\int_z^\infty dx
\frac{F_Z(\alpha,x;\eps)}{x^{1+\eps}}.\label{mod}\ee Here, \be Z=1+\sum_{k=1}^\infty \alpha^k p_k(\eps)\ee is assumed to be a series in $\alpha$ with
coefficients which are Laurent series in the regulator $\eps$ with poles of finite order and we thus set $p_k(\eps)=\sum_{j=-k}^\infty p_{k,j}\eps^j$ for some
real numbers $p_{k,j}$. A glance at Eq.(\ref{mod}) shows that the integrals involved in solving it as a fixpoint equation in $\alpha$ are all logarithmically
divergent at the upper boundary for $\eps=0$. All these integrals will indeed give Laurent series in $\eps$ with poles of finite order. Hence we attempt  to
choose the $p_k(\eps)$ such that the limit $\eps\to 0$ exists in Eq.(\ref{mod}). We want to understand the remaining ambiguity in that choice.

Let us first define the residue of our DSE as the pole at $\eps=0$ associated to the integral operator $\wp$ involved in it: \be {\rm Res}(\wp)=\lim_{\eps\to
0}\eps \int_z^\infty \frac{1}{x^{1+\eps}}dx.\ee Equally well  ${\rm Res}$ can be defined as the coefficient of the logarithmic growth at plus infinity of the
integral operator underlying our DSE: \be {\rm Res}(\wp)=-\lim_{\Lambda\to\infty}\frac{\alpha\int_z^\Lambda \frac{1}{x}dx}{\ln(\Lambda)}.\ee So by residue we
mean the coefficient of $\ln(z)$ in this integral, and hence it is closely related to the anomalous dimension $\gamma(\alpha)$ of our Green function, defined as
the coefficient of logarithmic growth with respect to a dimensionful variable, \be\gamma(\alpha)=\partial_{\ln(z)}\ln[F(\alpha,z;\eps)]|_{\eps=0,z=1}.\ee

This is in accordance with the operator-theoretic residue to which this generalizes in the case of Feynman graphs considering the primitive elements of their
corresponding Hopf algebra. In the models in subsequent sections below we will see that in general the function $\gamma(\alpha)$ is not merely given by the
residue at the primitive element $t_1$ as will be the case in this section, though the residue continues to play the most crucial role in the determination of an
anomalous dimension. Here, for our DSE above,  ${\rm Res}(\wp)=1$.

Regard (\ref{mod}) as a fixpoint equation for $F_Z$ and set \be F_Z=Z+\sum_{k=1}^\infty \alpha^k c_k^Z(z;\eps).\label{mode}\ee The notation emphasizes the
dependence on the "counterterm" $Z$. Let us first set $Z=1$ in (\ref{mod}), ie.~$p_k(\eps)=0$ $\forall k$. We regard Eq.(\ref{mod}) as an unrenormalized DSE for
the Hopf algebra of ladder trees, with  Feynman rules exemplified shortly.

We find, plugging (\ref{mode}) in (\ref{mod}), \bea c_1^{Z=1}(z;\eps) &  = & \int_z^\infty dx
\frac{x^{-\eps}}{x}=z^{-\eps}\frac{1}{\eps},\\
c_2^{Z=1}(z;\eps) & = & \frac{z^{-2\eps}}{2!\eps^2},\eea and in general \be c_k^{Z=1}(z;\eps)  =  z^{-k\eps}\frac{1}{k!\eps^k}.\ee

Let us set \be c_k^{Z=1}(z;\eps)=\sum_{j=0}^k c_{k,j}^{Z=1}(\eps)\ln^j(z),\ee upon expanding $z^{-\eps}$ (discarding terms $\ln(z)^j$ with $j>k$ as they will
always drop out ultimately when $\eps\to 0$ as the powers of $\ln(z)$ are always bounded by the augmentation degree), heading towards the two gradings in
$\alpha$ and $\ln(z)$. The coefficients $c_{k,j}^{Z=1}$ are Laurent series in $\eps$ with poles of finite order as promised. Actually, we see that they are
extremely simple in this first example. This will change soon enough, and certainly does in full QFT.

Before we solve our DSE exactly, let us set up the perturbative approach in analogy to perturbative quantum field theory. We use the ladder trees $t_n$ as
elements of the Hopf algebra $H$ and with multiplication $t_n\cdot t_m$, so that \be \Delta(t_n\cdot t_m )=\Delta(t_n)\cdot \Delta(t_m),\ee where $(h_1\otimes
h_2)\cdot (h_3\otimes h_4)=h_1\cdot h_3\otimes h_2\cdot h_4$.

Again, define Feynman rules $\phi$ this time by \be \phi\left(B_+(h)\right)(z;\eps)= \int_z^\infty dx \frac{\phi\left[h\right](x;\eps)}{x}, \ee and
$\phi(1)(z;\eps)=1$ $ \forall z,\eps$. With such Feynman rules we immediately have
\begin{prop}
\be c_k^{Z=1}(z;\eps)=\phi(t_k)(z;\eps).\ee
\end{prop}
This allows to regard Eq.(\ref{mod}) as the image under those Feynman rules $\phi$ of the already familiar combinatorial fix-point equation \be X=1+\alpha
B_+(X).\ee  As a side remark, we note that \be\phi(t_n\cdot
t_m)(z;\eps)=\frac{(n+m)!}{n!m!}\phi(t_{n+m})(z;\eps)=\frac{z^{-(n+m)\eps}}{n!m!\eps^{n+m}}=\phi(t_n)(z;\eps)\phi(t_m)(z;\eps).\ee This factorization of the
Feynman rules even on a perturbative level is a property of the simplicity of this first model. It holds in general in any renormalizable quantum field theory
for the leading pole term, as can be easily shown in any complex regularization like dimensional regularization or analytic regularization, for that matter
\cite{Chen,DK}.

Note that we have two different expansion parameters in our DSE. There is $\alpha$, but for each coefficient $c_k(z;\eps)$ we can expand this coefficient in
terms of powers of $\ln(z)$. As we are interested in the limit $\eps\to 0$, it is consistent to maintain only coefficients which have a pole or finite part in
$\eps$ as we did above. This gives a second grading which, in accord with quantum field theory \cite{annals}, is provided by the augmentation degree
\cite{annals,Kreimer2}. Note that this is consistent with what we did in the previous section, upon noticing that $-{\rm Li}_k(z)\sim \ln(z)^k/k!={\rm L}_k(z)$
for $\mid\!z\!\mid\to\infty$, so that indeed all rows had decreasing degree in $\ln(z)$ from right to left.

Hence we should feel tempted to organize the perturbative solution to our unrenormalized DSE in a manner using again a lower triangular matrix. This does not
look very encouraging for the unrenormalized solution though: let us set \be M_{i,j}^{(N)}=c^{Z=1}_{i,i-j}(\eps)\ln^{i-j}(z),\ee making use of both gradings.
Looking at this matrix for say $N=4$, we find  \be M^{(4)}=\left(\begin{array}{rcrcrcr}
1 & \mid & 0 & \mid & 0 & \mid & 0\\
-{\rm L}_1(z) & \mid & \frac{1}{\eps} & \mid & 0 & \mid & 0\\
2{\rm L}_2(z) & \mid & -\frac{1}{\eps} {\rm L}_1(z) & \mid & +\frac{1}{2!\eps^2}
 & \mid & 0\\
 -\frac{9}{2}{\rm L}_3(z) & \mid & +\frac{3}{2\eps}{\rm L}_2(z) & \mid & -\frac{1}{2!\eps^2}{\rm L}_1(z) & \mid & +\frac{1}{3!\eps^3}\end{array}\right) \ee
where again orders in $\alpha$ increase top to bottom and orders in $\ln(z)$ from right to left. This matrix $M$ is an unrenormalized matrix, its evaluation at
$\eps=0$ is impossible. Worse, it does not reveal much structure similar to what we had previously. But so far, this matrix is completely meaningless, being
unrenormalized. Thus, being good physicists, our first instinct should be to renormalize it by local counterterms. This will lead us, as we will see, just back to
the desired structural properties.

To renormalize it, we have to choose  $Z\not= 1$ such that the poles in $\eps$ disappear, by choosing appropriate $p_k(\eps)=\sum_{j=-k}^\infty p_{k,j}\eps^j$.
To understand the possible choices let us go back to the simple case $N=2$ (i.e.~calculating to order $\alpha$ merely) for which we obtain for a generic
$Z=1+\alpha p_1(\eps)$ \be \left(
\begin{array}{rcr} 1 & \mid & 0\\ -\ln z & \mid & \frac{1}{\eps}+p_1(\eps)\end{array}\right)\ee
As we require that $M^{(2)}(z;\eps)$ exists at $\eps=0$, this fixes $p_{1,-1}$: \be \left\langle\frac{1}{\eps}+p_1(\eps)\right\rangle=0\rightarrow
p_{1,-1}=-\frac{1}{\eps},\ee where $\langle\ldots\rangle$ means projection onto the pole part. All higher coefficients $p_{1,j}$, for $j=0,1,\ldots$, are left
undetermined. To understand better the full freedom in that choice of a renormalized $M^{(N)}$, let us reconsider perturbative renormalization for $M^{(N)}$. It
is indeed clear that we are confronted with a choice here: we absorb singularities located at $\eps=0$ and hence there is a freedom to choose the remaining
finite part. In physicists parlance this corresponds to the choice of a renormalization scheme. But such maps can not be chosen completely arbitrarily: they must
be in accord with the group structure of the character group of the Hopf algebra, and they must leave the short-distance singularities untouched. Both
requirements are easily formulated. For the first, we introduce a Rota--Baxter map $R$ \cite{Chen,int2}, \be R[ab]+R[a]R[b]=R[R[a]b]+R[aR[b]].\label{req1}\ee For
the second we demand that it is chosen such that \be R\left[c_k^{Z=1}(z_0;\eps)\right]-c_k^{Z=1}(z_0;\eps)\label{req2}\ee exists at $\eps=0$ for all $k$: at a
given reference point $z_0$, usually called the renormalization point, we require that the Rota--Baxter map leaves the short-distance singularities reflected in
the poles in $\eps$ unaltered. From now on we shall set the renormalization point to $z_0=1$ for simplicity.

Define the Bogoliubov map with respect to $R$, $\overline{\phi}_R$,  by \be \overline{\phi}_R(t_n)=m (S_R^\phi\otimes\phi)({\rm id}\otimes P)\Delta(t_n),\ee with
$P$ still the projector into the augmentation ideal. Note that indeed we had this equation before in (\ref{prop1}).

Now we have a Birkhoff decomposition of the Feynman character $\phi$ with respect to $R$ \be\phi_+=[{\rm
id}-R](\overline{\phi}_R),\;\phi_-=-R(\overline{\phi}_R),\ee for any Rota--Baxter map as above \cite{Chen}, into the renormalized character $\phi_+$ and the
counterterm $\phi_-$, thanks to the existence of a double construction which brings renormalization close to integrable systems for any renormalization scheme
$R$ \cite{int2}. The crucial fact here is that the pole parts which are still present in the Bogoliubov map are free of $\ln(z)$, which makes sure that $\phi_-$
provides local counterterms:
\begin{theorem}
$\lim_{\eps\to 0}\frac{\partial}{\partial\ln(z)}\overline{\phi}_R(t_n)(z;\eps)$ exists for all $n$.
\end{theorem}
Proof: The theorem has been proven much more generally \cite{annals,BK}. A proof follows immediately from induction over the augmentation degree, using that \be
S_R^\phi(B_+(t_n))=-R[m\circ(S_R^\phi\otimes \phi)\circ({\rm id}\otimes B_+)\Delta(t_n)],\ee using the Hochschild closedness $bB_+=0$ and the fact that each
element  in the perturbation series is in the image of such a closed 1-cocycle. This connection between Hochschild closedness and locality is universal in
quantum field theory \cite{Kreimer2,BK}, and will be discussed in detail in \cite{new}.

\medskip

Let us look at an example. \bea \lim_{\eps\to 0}\frac{\partial}{\partial \ln(z)}\overline{\phi}_R(t_2)(z;\eps) & = & \lim_{\eps\to 0}\frac{\partial}{\partial
\ln(z)}\left(\frac{1}{2!\eps^2}z^{-2\eps}-\left(\frac{1}{\eps}+p_{1,0}\right)\frac{1}{\eps}z^{-\eps}\right)\\ & = & p_{1,0}+\ln(z),\nonumber\eea where $p_{1,0}$
depends on the chosen renormalization scheme $R$.

So this theorem tells us that the pole terms in $\overline{\phi}_R$ are local, independent of $\ln(z)$. Now, every choice of $R$ as above determines a possible
$Z$ in the DSE by setting \be Z=1+\sum_{n=1}^\infty \alpha^n S_R^\phi(t_n).\ee We can hence introduce the renormalized matrix $M_{i,j}^{(N),R}(z,\eps)$ for any
such $R$. In particular, we can consider this matrix for the renormalized character \be[{\rm id}-R](\overline{\phi}_R)= S_R^\phi\star\phi(X)\equiv
m(S_R^\phi\otimes\phi)\Delta,\ee so that $\star$ denotes the group law in the character group of the Hopf algebra. The above proposition then guarantees that the
corresponding matrix exists at $\eps=0$, by the choice of $\ln(z)$-independent $p_k(\eps)$.

Renormalization has achieved our goal. Now the renormalized matrix $M^{(N),R}(z,0)$ has the same structure as before: columnwise, the coefficient of a power of
$\ln(z)$ is inherited from the row above. Let us look at $M^{(4),R}$ chosing a renormalized character $\phi_+$ with $R$ chosen to be evaluation at $z=1$, which in
this simple model agrees with the projection onto the pole part so that subtraction at the renormalization point is a minimal subtraction (MS) scheme (as
$\phi(t_n)\sim\frac{z^{-n\eps}}{n!\eps^n}$ only has poles and no finite parts in $\eps$). \be \left(\begin{array}{rcrcrcr}
  1 & \mid & 0 & \mid & 0 & \mid & 0\\
  -{\rm L}_1(z) & \mid  & 0 & \mid & 0 & \mid  & 0\\
  +{\rm L}_2(z) & \mid & 0 & \mid & 0 & \mid & 0\\
  -{\rm L}_3(z) & \mid & 0 & \mid & 0 & \mid & 0
\end{array}\right)\label{simple}\ee which is so simple
for this choice of $R$ that almost no structure remains. We nevertheless urge the reader to work $S_R^\phi(t_n)$ out  for several $n$ as in \be
S_R^\phi(t_2)=-R[\phi(t_2)+S_R^\phi(t_1)\phi(t_1)]=-R\left[\frac{1}{2!\eps^2}z^{-2\eps}\right]+R\left[R\left[
\frac{1}{\eps}z^{-\eps}\right]\frac{1}{\eps}z^{-\eps}\right]=\frac{1}{2\eps^2}.\ee

Due to the simplicity of this DSE we can now show that its perturbative solution in this MS scheme agrees with the non-perturbative (NP) solution for the same
renormalization point: at $z=1$, we require $F(\alpha,z)=1$. We immediately find that this leads to a Dyson--Schwinger equation \be F^{\rm NP}(\alpha,a)=1+\alpha
\left( \int_z^\infty dx \frac{F^{\rm NP}(\alpha,x)}{x} - \int_1^\infty dx \frac{F^{\rm NP}(\alpha,x)}{x} \right).\ee This reproduces the result Eq.(\ref{simple})
above. This agreement between the Taylor expansion of the non-perturbative solution and the renormalized solution in the MS scheme is a degeneracy of this simple
model.

We obviously have \be\phi^{\rm NP}(\alpha,z)=z^{-\alpha}\ee and \be \phi^{\rm NP}(t_{m+n})=\phi^{\rm NP}(t_m)\phi^{\rm NP}(t_n),\ee a hallmark of a
non-perturbative approach not available for a perturbative scheme, in particular not for a MS scheme.

 Note that we obtain scaling behaviour: $F(\alpha;z)=z^{-\alpha}$, thanks to the
basic fact that the DSE was linear. Indeed, the {\it Ansatz} $F(\alpha,z)=z^{-\gamma(\alpha)}$ solves the DSE above immediately as \be z^{-\gamma(\alpha)}=
1+\alpha\frac{{\rm Res}(\wp)}{\gamma(\alpha)}\left( z^{-\gamma(\alpha)}-1\right)\Leftrightarrow 1=\frac{\alpha{\rm Res(\wp)}}{\gamma(\alpha)},\ee delivering
$\gamma(\alpha)=\alpha$, as ${\rm Res}(\wp)=1$. Note that \be {\rm Res}(\wp) ={\rm Res_{\eps=0}}(\phi(t_1))=\lim_{\eps\to 0}\eps \phi(t_1)(z;\eps),\ee the
residue of the primitive element of the Hopf algebra, evaluated under the Feynman rules. This holds in general: at a conformal point (a non-trivial fixpoint of
the renormalization group) of a QFT one is to find scaling in a DSE and the anomalous dimension is just the sum of the residues of the primitive elements of the
Hopf algebra underlying the DSE.

It is high time to come back to the question about the freedom in chosing $R$. The simple Rota--Baxter map $R$ considered above led to Laurent polynomials
$p_k(\eps)$ which were extremely simple, in particular, $p_{k,j}$ was zero for $j\geq k$. Assume you make other choices, such that the requirements on $R$,
Eqs.(\ref{req1},\ref{req2}), are still fulfilled. In general, for such a generic $R$, we find here a solution \be F^R(\alpha,z)=(\tilde{z})^{-\alpha},\ee where
\be \tilde{z}=z\exp\{\Upsilon(\alpha)\} \equiv z\exp\left\{\sum_{j=0}^\infty \frac{\upsilon_j\alpha^j}{(j+1)!}\right\},\ee for coefficients $\upsilon_j$
recursively determined by the choice of $R$ (or $p_{j,k}$, respectively), and for example $M^{(4),R}$ looks like \be \left(\begin{array}{rcrcrcr}
  1 & \mid & 0 & \mid & 0 & \mid & 0\\
  -{\rm L}_1(z) & \mid  & -\upsilon_0 & \mid & 0 & \mid  & 0\\
  +{\rm L}_2(z) & \mid & +\upsilon_0{\rm L}_1(z) & \mid & +\frac{1}{2}(\upsilon_0^2-\upsilon_1) & \mid & 0\\
  -{\rm L}_3(z) & \mid & -\upsilon_0{\rm L}_2(z) & \mid & -\frac{1}{2}\left(\upsilon_0^2-\upsilon_1\right){\rm L}_1(z) & \mid &
  -\frac{1}{3!}\left(\upsilon_0^3+3\upsilon_0\upsilon_1-\upsilon_2\right)
\end{array}\right)
\ee Note that the associated DSE has the form \be F_R(\alpha,z)=F_R(\alpha,1)-\alpha\int_z^1\frac{F_R(\alpha,x)}{x},\ee where
$F_R(\alpha,1)=\exp\left\{\sum_{j=0}^\infty \upsilon_j\alpha^j\right\}$.

So finally, the ambiguities in the choice of a finite part in renormalization and in the choice of a branch for the log are closely related, a fact which is
similarly familiar in quantum field theory in the disguise of the optical theorem connecting real and imaginary parts of quantum field theory amplitudes, as will
be discussed elsewhere. Finally, we note that the solution to our DSE fulfills \be \frac{\partial \ln F_R(\alpha,z)}{\partial \ln(z)}=-\alpha\ee for all $R$,
confirming the renormalization scheme independence of the anomalous dimension $\gamma_F(\alpha)=-\alpha{\rm Res}(\wp)$ of the Green function $F_R(\alpha,z)$, a
fact which generally holds when dealing with a DSE which is linear. This last equation actually is a remnant of the propagator coupling duality in quantum field
theory, first explored in \cite{BrK}.
\subsection{Another toy: $F(\alpha,z)=z^{\arcsin[\alpha\pi{\rm Res}(\wp)]/\pi}$}
Next, let us study yet another DSE, which is slightly more interesting in so far as that the anomalous dimension is not just given by the residue of the integral
operator on the rhs of the equation. Consider the DSE \be F(\alpha,z;\eps)=Z+\alpha\int_0^\infty \frac{F(\alpha,x;\eps)}{x+z}dx. \ee First note that again ${\rm
Res}(\wp)=1$. Continuing, we find \bea c_1^{Z=1}(z;\eps) &  = & \int_0^\infty dx
\frac{x^{-\eps}}{x+z}=z^{-\eps}\frac{1}{\eps}B(1-\eps,1+\eps),\\
c_2^{Z=1}(z;\eps) & = & \frac{z^{-2\eps}}{2!\eps^2}B_1B_2, \eea and in general \be c_k^{Z=1}(z;\eps)  =  z^{-k\eps}\frac{1}{k!\eps^k}B_1\ldots B_k,\ee where
$B_k:=B(1-k\eps,1+k\eps)$.

As before, let us set \be c_k^{Z=1}(z;\eps)=\sum_{j=0}^k c_{k,j}^{Z=1}(\eps)\ln^j(z),\ee upon expanding $z^{-\eps}$. The coefficients $c_{k,j}^{Z=1}$ are again
Laurent series in $\eps$ with poles of finite order. In this example we can indeed distinguish between the perturbative solution in the MS scheme and the
non-perturbative solution of the DSE, as $c_k(1,\eps)$ is a Laurent series in $\eps$ which has non-vanishing finite and higher order parts. It has some merit to
study both the MS and the NP case. In the MS scheme we define $R$ to evaluate at the renormalization point $z=1$ and to project onto the proper pole part. This
defines indeed a Rota--Baxter map \cite{Chen,int2}, and as an example, let us calculate \bea \overline{\phi}_{\rm MS}(t_2) & = &
\frac{1}{2!\eps^2}B_1B_2z^{-2\eps}-\frac{1}{\eps^2}B_1z^{-\eps},\nonumber\\
 & = & -\frac{1}{2\eps^2}-\frac{3}{2}\zeta(2)+{\rm L}_2(z),\eea disregarding terms which vanish at $\eps=0$. In accordance with our theorem, no pole
 terms involve powers of $\ln(z)$.
 The counterterm $S_{\rm MS}^\phi(t_2)$ subtracts these pole terms only, leaving $\phi_+(t_2)={\rm L}_2(z)-\frac{3}{2}{\rm Li}_2(1)$, where $\zeta(2)={\rm Li}_2(1)$.

For the MS scheme we hence find a renormalized matrix \be M^{(4),{\rm MS}}(z,0) = \left(\begin{array}{rcrcrcr}
  1 & \mid & 0 & \mid & 0 & \mid & 0\\
  -{\rm L}_1(z) & \mid & 0 & \mid & 0 & \mid & 0\\
   +{\rm L}_2(z) & \mid & 0 & \mid & -\frac{3}{2}{\rm Li}_2(1) & \mid & 0\\
   -{\rm L}_3(z) & \mid & 0 & \mid & +\frac{3}{2}{\rm L}_1(z){\rm Li}_2(1) & \mid & 0\\
\end{array}\right)
\ee Note that this still has non-zero entries along the diagonal, so that $F_{\rm MS}(\alpha,1)=1+{\mathcal O}(\alpha)$.

 Non-perturbatively, we find a solution by imposing the  side constraint $F(\alpha,1)=1$ as \be
F_{\rm NP}(\alpha,z)=z^{\arcsin[\alpha\pi{\rm Res}(\wp)]/\pi}.\ee Note that now the corresponding entries in the Matrix $M^{(N),\rm NP}_{i,j}$ are not only located
in the leftmost column, but are given by the double Taylor expansion  \be M^{(N),\rm NP}_{i,j}=\frac{\partial_{\alpha}^i\partial_{\ln(z)}^j}{i!j!}
\exp\left\{\frac{\arcsin(\pi\alpha)}{\pi}\ln(z)\right\}_{\alpha=0,\ln(z)=0}.\ee The residue here is still simple: ${\rm Res}(\wp)=1$. Again, we can find the
above solution with the Ansatz (scaling) \be F(\alpha,z)=z^{-\gamma(\alpha)}\ee where we assume $\gamma(\alpha)$ to vanish at $\alpha=0$. With this Ansatz we
immediately transform the DSE into \be z^{-\gamma(\alpha)}-1=\alpha \frac{{\rm
Res}(\wp)}{\gamma(\alpha)}B(1-\gamma(\alpha),1+\gamma(\alpha))[z^{-\gamma(\alpha)}-1]\ee from which we conclude \be
\gamma(\alpha)=\frac{\arcsin\alpha\pi}{\pi}=\alpha+\zeta(2)\alpha^3+\cdots.\ee Note that this solution has branch cuts outside the perturbative region $\mid
\alpha\mid<1$. Furthermore, note that the same solution is obtained for the DSE \be F(\alpha,z)=\alpha\int\frac{F(\alpha,x)}{x+z}dx,\ee as the inhomogenous term
is an artefact of the perturbative expansion which is absorbed in the scaling behaviour. Finally we note that the appearance of scaling is again a consequence of
the linearity of this DSE, and if we were to consider a DSE like \be F(\alpha,z)=\int_0^\infty \frac{{\mathcal F}(F(\alpha,x))}{x+z}dx,\ee say, for ${\mathcal
F}$ some non-linear polynomial or series, then indeed we would not find scaling behaviour. An Ansatz of the form \be
F(\alpha,z)=z^{-\gamma(\alpha)}\sum_{k=0}^\infty c_k(\alpha)\ln^k(z),\ee is still feasible though, and leads back to the propagator-coupling dualities \cite{BrK}
explored elsewhere.

Having determined the non-perturbative solution here by the boundary condition $F_{\rm NP}($ $\alpha,1)=1$, other renormalization schemes can be expressed
through this solution as before introducing $\tilde{z}=z\exp(\Upsilon(\alpha))$ for a suitable series $\Upsilon(\alpha)$. The difficulty with perturbative
schemes like MS is simply that we do not know off-hand the corresponding boundary condition for a non-perturbative solution of such a scheme as the series
$\Upsilon(\alpha)$ has to be calculated itself perturbatively, and often is in itself highly divergent as an asymptotic series in $\alpha$. Even if one were able
to resum the perturbative coefficients of a minimal subtraction scheme, the solution so obtained will solve the DSE only with rather meaningless boundary
conditions which reflect the presence not only of instanton singularities but, worse, renormalon singularities in the initial asymptotic series. While it is
fascinating to import quantum field theory methods into number theory, which suggest to resum perturbation theory amplitudes of a MS scheme making use of the
Birkhoff decomposition of \cite{CK2} combined with progress thanks to Ramis and others in resumation of asymptotic series, as beautifully suggested recently \cite{CoMa}, the
problem is unfortunately much harder still for a renormalizable quantum field theory. We indeed have almost no handle outside perturbation theory on such
schemes, while on the other hand the NP solution of DSE with physical side-constraints like $F(\alpha,1)=1$ is amazingly straightforward and resums perturbation
theory naturally once one has recognized the role of the Hochschild closed 1-cocycles \cite{annals,BrK,Kreimer2}. Such an approach will be exhibited in detail in
\cite{new}. The idea then to reversely import number-theoretic methods into quantum field theory is to my mind very fruitful and needed to make progress at a
level beyond perturbation theory. It is here where in my mind the structures briefly summarized in section three shall ultimately be helpful to overcome these
difficulties and make the kinship between numbers and quantum fields even closer.

\subsection{More like QFT: $F(\alpha,q^2/\mu^2)=[q^2/\mu^2]^{(1-\sqrt{5-4\sqrt{1-2\alpha {\rm Res}(\wp)}})}$}
Let us finish this section with one simple DSE originating in QFT, say a massless scalar field theory with cubic coupling in six dimensions, $\varphi^3_6$, with
its well-known Feynman rules \cite{CK1}. We consider the vertex function at zero-momentum transfer which obeys the following DSE (in a NP scheme such that
$F(\alpha,1)=1$) \be F\left(\alpha,\frac{q^2}{\mu^2}\right)=1+\alpha \int d^6k \frac{F(\alpha,\frac{k^2}{\mu^2})}{[k^2]^2[(k+q)^2]}.\ee The scaling Ansatz \be
F\left(\alpha,\frac{q^2}{\mu^2}\right)=\left(\frac{q^2}{\mu^2}\right)^{-\gamma(\alpha)}\ee still works which is rather typical \cite{BrK,Delb} and delivers \be
1=\frac{\alpha{\rm
Res}(\wp)}{\gamma(\alpha)(1-\gamma(\alpha))(1+\gamma(\alpha))(1-\gamma(\alpha)/2)}\Rightarrow\gamma(\alpha)=(1-\sqrt{5-4\sqrt{1-\alpha}})\label{conf}\ee using
that the residue is still very simple: \be {\rm Res}(\wp)=\frac{1}{2}.\ee We used that \be \int d^6k
\frac{\left(\frac{k^2}{\mu^2}\right)^{-x}}{[k^2]^{2}(k+q)^2}={\rm Res}(\wp) \left[\frac{q^2}{\mu^2}\right]^{-x}\frac{1}{x(1-x)(1+x)(1-x/2)},\ee which is
elementary. Note the invariance under the transformation $\gamma(\alpha)\to 1-\gamma(\alpha)$ in the denominator polynomial of Eq.(\ref{conf}) which reflects the
invariance of the primitive ${\rm Res}(\wp)=\phi(t_1)$ under the conformal transformation in momentum space $k_\nu\to k_\nu/k^2$ at the renormalization point
$q^2=\mu^2$.

\section{The real thing}
%\begin{fmffile}{fmfgalintr}
And how does this fare in the real world of local interactions, mediated by quantum fields which asymptotically approximate free fields specified by covariant
wave equations and a Fourier decomposition into raising and lowering operators acting on a suitable state space? The following discussion was essentially given
already in \cite{Kreimer2} and is repeated here with special emphasis on the analogies pointed out in the previous two sections.

Considering DSEs in QFT, one usually obtains them as the quantum equations of motion of some Lagrangian field theory using some generating functional technology
in the path integral. DSEs for 1PI Green functions can all be written in the form \be \Gamma^{\underline{n}}= 1 + \sum_{\genfrac{}{}{0pt}{}{\gamma\in
H_L^{[1]}}{{\rm res}(\gamma)=\underline{n}}} \frac{\alpha^{\vert\gamma\vert}}{{\rm Sym}(\gamma)} B_+^\gamma(X_{\mathcal R}^\gamma), \ee where the $B_+^\gamma$
are Hochschild closed 1-cocycles of the Hopf algebra of Feynman graphs indexed by Hopf algebra primitives $\gamma$ which are linear generators of the Hopf
algebra, and as primitives have augmentation degree 1, with external legs $\underline{n}$, and $X_{\mathcal R}^\gamma$ is a monomial in superficially divergent
Green functions which dress the internal vertices and edges of $\gamma$ \cite{Kreimer2,new}. This allows to obtain the quantum equations of motion, the DSEs for
1PI Green functions, without any reference to actions, Lagrangians or path integrals, but merely from the representation theory of the Poincar\'e group for free
fields.

Hence we were justified in this paper to call any equation of the form (and we only considered the linear case $k=1$ in some detail, while in general a
polynomial or even a series in $X$ can appear) \be X=1+\alpha B_+(X^k),\ee with $B_+$ a closed Hochschild 1-cocycle, a Dyson Schwinger equation. In general, this
motivates an approach to quantum field theory which is utterly based on the Hopf and Lie algebra structures of graphs \cite{annals}.
\subsection{Determination of $H$}
The first step aims at finding the Hopf algebra suitable for the description of a  chosen renormalizable QFT. For such a QFT, identify the one-particle
irreducible (1PI) diagrams. Identify all edges and propagators in them and define a pre-Lie product on 1PI graphs by using the possibility to replace a local
vertex by a vertex correction graph, or, for internal edges, by replacing a free propagator by a self-energy. For any local QFT this defines a pre-Lie algebra of
graph insertions \cite{annals}. For a renormalizable theory, the corresponding Lie algebra will be non-trivial for only a finite number of types of 1PI graphs
(self-energies, vertex-corrections) corresponding to the superficially divergent graphs, while the superficially convergent ones provide a semi-direct product
with a trivial abelian factor \cite{CK2}.

The combinatorial graded pre-Lie algebra so obtained provides not only a Lie-algebra ${\mathcal L}$, but a commutative graded Hopf algebra $H$ as the dual of its
universal enveloping algebra ${\mathcal U(L)}$, which is not cocommutative if ${\mathcal L}$ was non-abelian. Dually one hence obtains a commutative but
non-cocommutative Hopf algebra $H$ which underlies the forest formula of renormalization. This generalizes the examples discussed in the previous sections as
they were all cocommutative. The main structure, and the interplay between the gradings in $\alpha$ and $\ln(z)$ are maintained though, as a glance at
\cite{Kreimer2} easily confirms.
\subsection{Character of $H$} For such a Hopf algebra $H
= H (m,E,\bar e,\Delta ,S)$, a Hopf algebra with multiplication $m$, unit $e$ with unit map $E:{\mathbb Q}\to H$, $q\to qe$, with counit $\bar e$, coproduct
$\Delta$ and antipode $S$, $S^2 = e$, we immediately have at our disposal the group of characters $G= G(H)$ which are multiplicative maps from $G$ to some target
ring $V$. This group contains a distinguished element:
 the Feynman
rules $\varphi$ are indeed a very special  character in $G$. They will typically suffer from short-distance singularities, and the character $\varphi$ will
correspondingly reflect these singularities.  We will here typically take $V$ to be the ring of Laurent polynomials in some indeterminate $\eps$ with poles of
finite orders for each finite Hopf algebra element, and design Feynman rules so as to reproduce all salient features of QFT. The Feynman rules of the previous
sections were indeed a faithful model for such behaviour.

As $\varphi : H \rightarrow V$, with $V$ a ring, with multiplication $m_V$, we can introduce the group law \be \varphi \star \psi =  m_V \circ (\varphi \otimes
\psi) \circ \Delta \, , \ee and use it to define a new character \be S_R^{\phi} \star \phi \in G \, , \ee where $S_R^{\phi} \in G$ twists $\phi \circ S$ and
furnishes the counterterm of $\phi (\Gamma)$, $\forall \, \Gamma \in H$, while $ S_R^{\phi} \star \phi (\Gamma)$ corresponds to the renormalized contribution of
$\Gamma$. $S_R^\phi$ depends on the Feynman rules $\phi:\;H\to V$ and the chosen renormalization scheme $R:\; V\to V$. It is given by \be
S_R^\phi=-R\left[m_V\circ (S_R^\phi \otimes \phi)\circ ({\rm id}_H\otimes P)\circ\Delta\right]\; ,\ee where $R$ is supposed to be a Rota-Baxter operator in $V$,
and the projector into the augmentation ideal  $P:H\to H$ is given by $P={\rm id}-E\circ \bar{e}$.

The $\bar R$ operation of Bogoliubov is then given by \be  \bar \phi:= \left[m_V\circ (S_R^\phi \otimes \phi)\circ ({\rm id}_H\otimes P)\circ\Delta\right]\; ,\ee
and \be S_R^\phi\star\phi\equiv m_V\circ (S_R^\phi\otimes \phi)\circ\Delta=\bar\phi+S_R^\phi=({\rm id}_H-R)(\bar\phi )\ee is the renormalized contribution.
Again, this is in complete analogy with the study in the previous sections.
\subsection{Locality from $H$} The next step aims to show that locality of
counterterms is utterly determined by the Hochschild cohomology of Hopf algebras \cite{annals,BK}. Again, one can dispense of the existence of an underlying
Lagrangian and derive this crucial feature from the Hochschild cohomology of $H$.   What we are considering are spaces ${\mathcal H}^{(n)}$ of maps from the Hopf
algebra into its own $n$-fold tensor product, \be {\mathcal H}^{(n)}\ni \psi\Leftrightarrow \psi: H\to H^{\otimes n} \ee  and an operator \be b:\; {\mathcal
H}^{(n)}\to {\mathcal H}^{(n+1)}\ee  which squares to zero: $b^2=0$. We have for $\psi\in {\mathcal H}^{(1)}$ \be (b\psi)(a)=\psi(a)\otimes
e-\Delta(\psi(a))+({\rm id}_H\otimes \psi)\Delta(a)\ee  and in general \be (b\psi)(a)=(-1)^{n+1}\psi(a)\otimes e +\sum_{j=1}^n (-1)^j \Delta_{(j)}\left(
\psi(a)\right)+({\rm id}_H\otimes\psi)\Delta(a),\ee where $\Delta_{(l)}:H^{\otimes n}\to H^{\otimes (n+1)} $ applies the coproduct in the $j$-th slot of
$\psi(a)\in H^{\otimes n}$.

Locality of counterterms and finiteness of renormalized quantities follow indeed from the Hoch\-schild properties of $H$: the Feynman graph is in the image of a
closed Hochschild 1-cocycle $B_+^{\gamma}$, $b \, B_+^{\gamma} =  0$, i.e. \be \Delta \circ B_+^{\gamma} (X) = B_+^{\gamma} (X) \otimes e + ({\rm id} \otimes
B_+^{\gamma}) \circ \Delta (X) \, , \ee and this equation suffices to prove the above properties by a recursion over the augmentation degree of $H$, again in
analogy to the study in the previous section.

\subsection{Combinatorial DSEs from Hochschild cohomology}
Having understood the mechanism which achieves locality step by step in the perturbative expansion, one realizes that this mechanism delivers the quantum
equations of motion, our DSEs. Once more, they typically are of the form \be \Gamma^{\underline{n}}= 1 + \sum_{\genfrac{}{}{0pt}{}{\gamma\in H_L^{[1]}}{{\rm
res}(\gamma)=\underline{n}}} \frac{\alpha^{\vert\gamma\vert}}{{\rm Sym}(\gamma)} B_+^\gamma(X_{\mathcal R}^\gamma)=1+\sum_{\genfrac{}{}{0pt}{}{\Gamma\in H_L}{
{\rm res}(\Gamma)=\underline{n}}}\frac{\alpha^{\vert\Gamma\vert}\Gamma}{{\rm Sym}(\Gamma)}\; , \ee
 where the first sum is over a finite (or
countable) set of Hopf algebra primitives $\gamma$, \be \Delta(\gamma)=\gamma\otimes e + e \otimes \gamma,\ee  indexing the closed Hochschild 1-cocycles
$B_+^{\gamma}$ above, while the second sum is over all one-particle irreducible graphs contributing to the desired Green function, all weighted by their symmetry
factors. Here, $\Gamma^{\underline{n}}$ is to be regarded as
 a formal series
\be  \Gamma^{\underline{n}}=1+\sum_{k\geq 1} c_k^{\underline{n}} \alpha^k, \;c_k^{\underline{n}}\in H.\ee These coefficients of the perturbative expansion
deliver sub-Hopf algebras in their own right \cite{Kreimer2}.

There is a very powerful structure behind  the above decomposition into Hopf algebra primitives - the fact that the sum over all Green functions
$G^{\underline{n}}$ is indeed the sum over all 1PI graphs, and this sum, the effective action,  gets a very nice structure: $\prod\frac{1}{1-\gamma}$, a product
over "prime" graphs - graphs which are primitive elements of the Hopf algebra and which index the closed Hochschild 1-ccocycles, in complete factorization of the
action. A single such Euler factor with its corresponding DSE and Feynman rules was evaluated in \cite{BrK}, a calculation which was entirely based on a
generalization of our study: an understanding of the weight of contributions $\sim \ln(z)$ from a knowledge of the weight of such contributions of lesser degree
in $\alpha$, dubbed propagator-coupling duality in \cite{BrK}. Altogether, this allows to summarize the structure in QFT as a vast generalization of the
introductory study in the previous sections. It turns out that even the quantum structure of gauge theories can be understood along these lines \cite{Kreimer2}.
A full discussion is upcoming \cite{new}.

Let us finish this paper by a discussion of the role of matrices $M^{(\gamma)}$ which one can set up for any Hochschild closed 1-cocycle $B_+^\gamma$ in the Hopf
algebra. The above factorization indeed allows to gain a great deal of insight into QFT from studying these matrices separately, disentangling DSEs into one
equation for each of them, of the form \be F^{(\gamma)}(\alpha,z)=1+\alpha^{\mid\!\gamma\!\mid}\int {\mathcal D}(\gamma, F^{(\gamma)}(\alpha,k))dk,\ee where
${\mathcal D}(\gamma, F^{(\gamma)}(\alpha,k))$ is the integrand for the primitive, which determines a residue which typically and  fascinatingly is not a boring
number $1,1/2,\ldots$ as in our previous examples, but a multiple zeta value in its own right \cite{annals}. Those are the numbers we should know and understand
for the benefit of quantum field theory - know them as motives and understand the contribution of their DSE to the full non-perturbative theory.

The above gives a linear DSE whose solution can be obtained by a scaling Ansatz as before. This determines an equation \be
1=\alpha^{\mid\!\gamma\!\mid}J_\gamma({\rm anom}_\gamma(\alpha))\ee leading to a dedicated anomalous dimension ${\rm anom}_\gamma(\alpha)$, just as we did
before, with $J_\gamma$ an algebraic or transcendental function as to yet only known in very few examples. Realizing that the breaking of scaling is parametrized
by insertions of logs into the integrand ${\mathcal D}$ with weights prescribed by the $\beta$-function of the theory one indeed finds a vast but fascinating
generalization of the considerations before.

In particular, matrices $M^{(\gamma)}$ can be obtained from a systematic study of the action of operators $S\star Y^k$, where $Y$ is the grading wrt to the
augmentation degree, which faithfully project onto the coefficients of $\ln^k(z)$ apparent in the expansion of $\ln F^{(\gamma)}(\alpha,z)$, as in \cite{BrK}. In
our previous examples this was simply reflected by the fact that $S\star Y^k(t_m)=0$ for $m>k$, so that for example the coefficient of $\ln(z)$ was only given by
the residue of $\phi(t_1)$, which upon exponentiation delivers the subdiagonal entries $M_{j+1,j}$, and similarly $S\star Y^k$ delivers the subdiagonals
$M_{j+k,j}$. To work these matrices out for primitives $\gamma$ beyond one loop (essentially, \cite{BrK} did it for one-loop) is a highly non-trivial exercise in
QFT, with great potential though for progress in understanding of those renormalizable theories. Apart from the perturbative results well-published already, and
the introductory remarks here and in \cite{Kreimer2}, a detailed study of DSEs in QFT will be given in \cite{new}.

%\end{fmffile}
\section*{Acknowledgments}
Let me first thank Katia Consani, Yuri Manin and Mathilde Marcolli for making the workshop possible. Whilst in my talk I reviewed mainly the Hopf algebraic
approach to renormalization in more detail than section three above provides, the somewhat polylogarithmically enthused view presented here is a result of
stimulating discussions with Spencer Bloch and Herbert Gangl, which I very gratefully acknowledge. And thanks again to Herbert for proofreading the ms.

\end{document}